\documentclass[journal=ancac3,manuscript=article,layout=twocolumn]{achemso}

\usepackage[T1]{fontenc}
\usepackage{graphicx}
\usepackage{amsmath}

\newcommand{\MoSS}{MoS$_2$$-$S}
\newcommand{\MoSeSe}{MoSe$_2$$-$Se}
\newcommand{\JanS}{MoSSe$-$S}
\newcommand{\JanSe}{MoSSe$-$Se}
\newcommand{\fmig}{f_{\mathrm{mig}}}
\newcommand{\fcr}{f_{\mathrm{cr}}}

\author{Bruno Ipaves}
\affiliation{Department of Applied Physics, Universidade Estadual de Campinas (UNICAMP), Campinas, 13083-859, S\~ao Paulo, Brazil}
\author{Guilherme S. L. Fabris}
\affiliation{Department of Applied Physics, Universidade Estadual de Campinas (UNICAMP), Campinas, 13083-859, S\~ao Paulo, Brazil}
\author{Raphael B. Oliveira}
\affiliation{Institute of Physics,
Federal University of Rio Grande do Norte, Natal, 59078-970, Rio Grande do Norte, Brazil}
\author{\\Pulickel M. Ajayan}
\affiliation{Department of Materials Science and NanoEngineering, Rice University, Houston, Texas 77005, United States}
\author{Marcelo L. Pereira Junior}
\email{marcelo.lopes@unb.br}
\affiliation{Department of Electrical Engineering, College of Technology, University of Bras\'ilia, 70910-900, Bras\'ilia, Federal District, Brazil}
\author{Douglas S. Galv\~ao}
\email{galvao@ifi.unicamp.br}
\affiliation{Department of Applied Physics, Universidade Estadual de Campinas (UNICAMP), Campinas, 13083-859, S\~ao Paulo, Brazil}

\title{On the Spontaneous Structural Reconstructions of Chalcogen-Depleted MoS$_2$, MoSe$_2$, and Janus MoSSe Monolayers}


\begin{document}



\begin{abstract}
Removing one chalcogen layer from a transition-metal dichalcogenide monolayer leaves an undercoordinated sheet that reconstructs as the remaining chalcogen atoms cross the metal plane. Here, we combine \textit{ab initio} and reactive molecular dynamics (MD) simulations of depleted MoS$_2$, MoSe$_2$, and Janus MoSSe, examining each remaining chalcogen on two parent lattices, to show that the remaining chalcogen type determines the structural reconstruction, with only a minor influence from the parent lattice. In \textit{ab initio} trajectories of the Janus-derived sheets, Se crosses the Mo plane already at 300~K, whereas S stays in place at 300~K and the sheet reconstructs only at 900~K. With a free in-plane cell, the crossings precede an anisotropic lattice contraction, which therefore results from the structural reconstruction. Reactive MD simulations of 50~nm sheets show that, when S remains, the Mo lattice stays crystalline in nanometric domains that disappear with time and temperature, with pores/holes at 900~K. In contrast, when Se remains, the Mo lattice loses crystalline order within the first picosecond, and 36--39\% of the Se atoms cross at every temperature. In a depleted lateral MoS$_2$/MoSe$_2$ heterostructure, both modes coexist in one sheet, each half reconstructing like its single-chalcogen counterpart, suggesting chalcogen patterning as a route to define crystalline and disordered regions in a single monolayer.

\end{abstract}

\section{Introduction}

Two-dimensional transition-metal dichalcogenides (TMDs) have attracted considerable attention because their structural, mechanical, electronic, and optical properties can be tuned by composition, including Janus and alloyed phases, dimensionality, and defect engineering \cite{lu2017janus, apte2018telluride, ye2016defects, pereira2020elastic, abhijith2026quasi}. Chalcogen vacancies are among the most common defects in these monolayers \cite{zhou2013intrinsic}, and the undercoordinated metal atoms they leave affect a broad range of properties. They mediate the exchange of chalcogen species and alloy formation \cite{taghinejad2018defect}, modify adsorption, and can induce magnetism \cite{fabris2026magnetism, wadumesthri2026interfacial}, while creating catalytically active sites in otherwise inert basal planes \cite{ye2016defects, li2016activating, lunardon2023catalytic, shi2019triggering, zhang2017janus, camposmata2026probing}. More generally, defect formation, adsorption, and diffusion control the stability and functionality of low-dimensional TMD phases \cite{ipaves2026unraveling, fabris2026magnetism, lei2016surface}.

Janus TMD monolayers are an appealing platform in this context. Their chemically distinct chalcogen terminations break out-of-plane mirror symmetry, producing an intrinsic dipole and responses absent in symmetric TMDs \cite{lu2017janus, zhang2017janus}; calculations also indicate that the two faces interact differently with an adjacent AlN sheet \cite{santos2021predicting}. Janus MoSSe has been obtained by replacing one chalcogen layer of a binary monolayer, either by stripping the top S layer of MoS$_2$ with a hydrogen plasma, which leaves the hydrogen-covered intermediate MoSH, followed by selenization \cite{lu2017janus, wan2021synthesis}, or by the direct sulfurization of the top Se layer of MoSe$_2$ \cite{zhang2017janus}. Atomic-layer substitution has also been achieved by selective replacement of the top chalcogen layer at room temperature \cite{guo2021designing}.

Experiments provide a direct view of how chalcogen-deficient monolayers rearrange as the chalcogen loss increases. Scanning transmission electron microscopy of monolayer MoS$_2$ has imaged its intrinsic point defects, dislocations, and grain boundaries \cite{zhou2013intrinsic}, and atomic-resolution electron microscopy has followed the diffusion pathways of Mo adatoms and vacancies \cite{hong2017direct}. In monolayer MoSe$_2$, Se vacancies mediate Se exchange with S \cite{taghinejad2018defect}, and Se deficiency is accommodated by mirror twin boundaries and inversion domains, in which the Mo sublattice itself rearranges \cite{lehtinen2015atomic, lin2015vacancy}. The outcome of strong chalcogen loss therefore depends on how the remaining atoms rearrange and how many atoms are removed. The limit of this sequence is the complete removal of one chalcogen layer, in which every Mo atom loses half of its bonds.

To our knowledge, the complete removal of one chalcogen layer has been simulated only for MoS$_2$. Reactive simulations of sulfur-depleted MoS$_2$ showed that the remaining S atoms cross the Mo plane and that the sheet evolves into a MoS$_x$ alloy with nanometric domains separated by line-like boundaries \cite{jaques2026structural}. Three points remain open. It is not known whether the remaining chalcogen changes how the depleted sheet reconstructs, whether the parent lattice (binary or Janus) has any influence, or whether the reconstructed sheet retains the crystalline order of the Mo lattice beyond the ~1.2~nm cell and the ~10~ps accessible to first-principles dynamics. These questions are especially relevant for the Janus phase, since each of its two faces can be removed and the two resulting sheets differ only in the chalcogen left.

In this work, we combine \textit{ab initio} molecular dynamics (AIMD) and reactive molecular dynamics to determine how the reconstruction of a chalcogen-depleted monolayer depends on the remaining chalcogen. We consider four monolayers obtained by removing one chalcogen layer from MoS$_2$, MoSe$_2$, and Janus MoSSe, which we call the parent and the removed layer: \MoSS, \MoSeSe, \JanS, and \JanSe, so each remaining chalcogen is examined on two different parent lattices. AIMD can identify the individual crossings of chalcogen atoms through the Mo plane and the lattice response. At the same time, reactive simulations of 50 $\times$ 50~nm sheets can capture the Mo lattice structure beyond the nanometer scale. We show that the remaining chalcogen defines the reconstruction mode, with the parent lattice playing a minor role. When S remains, the crossings are thermally activated, and the Mo lattice stays crystalline in nanometric domains. In contrast, when Se remains, the crossings occur already at room temperature, and the Mo lattice loses its crystalline order. In a depleted lateral MoS$_2$/MoSe$_2$ heterostructure, the two modes coexist in a single sheet, each confined to its own half.

\section{Results and Discussion}

\begin{figure*}[t!]
  \centering
  \includegraphics[width=16cm]{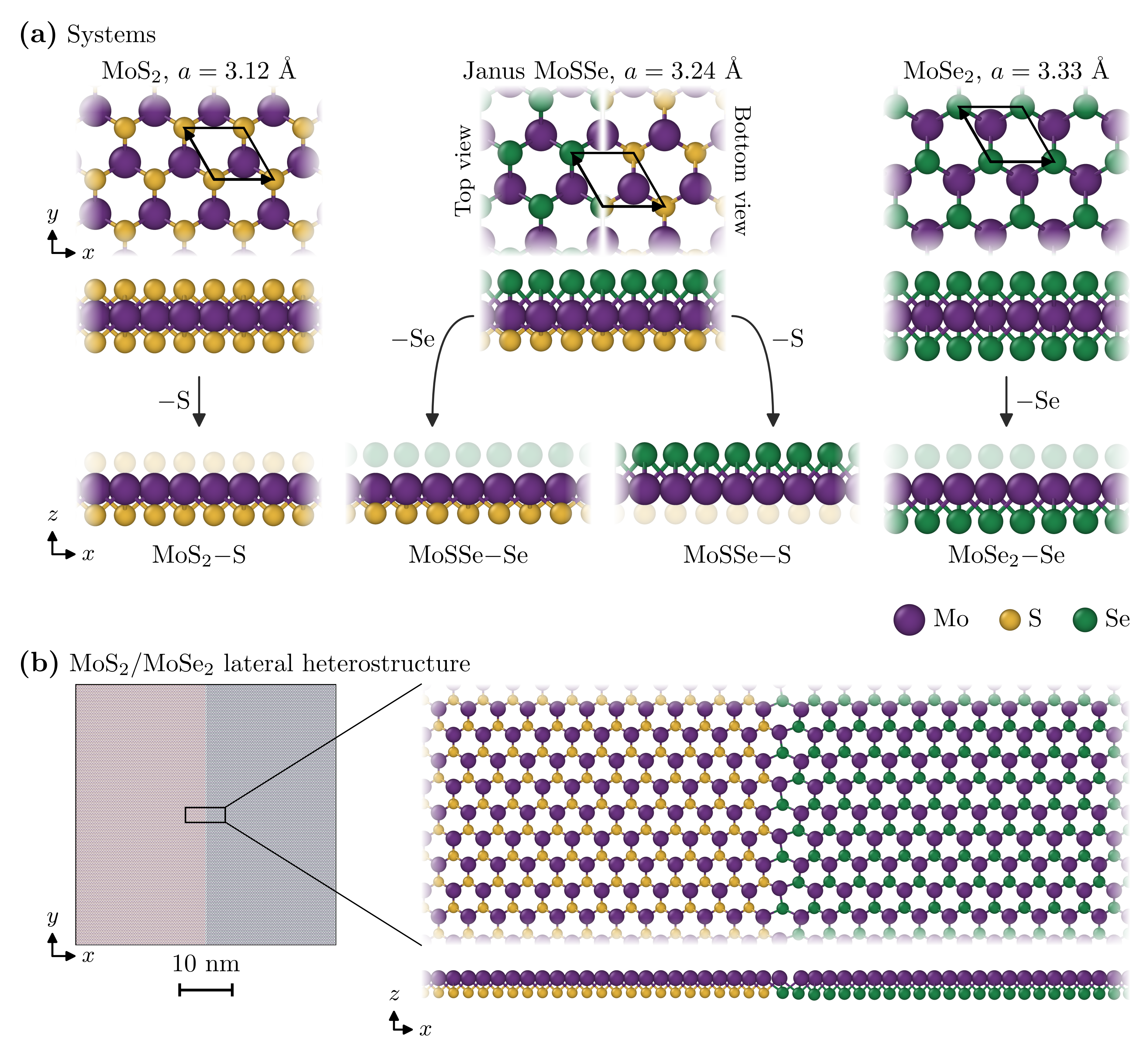}
  \caption{(a) Parent monolayers MoS$_2$, Janus MoSSe, and MoSe$_2$, with their primitive cells, and the four depleted monolayers obtained by removing one chalcogen layer, drawn as a virtual structure. (b) Depleted lateral MoS$_2$/MoSe$_2$ heterostructure used in the reactive simulations, with top and side views of the interface.
}
  \label{fig:systems}
\end{figure*}

We can separate the role of the remaining chalcogen from that of the parent lattice only if we examine each chalcogen on more than one parent. We therefore consider the three parent monolayers and the four depleted monolayers shown in Figure~\ref{fig:systems}(a), in which S remains in \MoSS{} and \JanSe{} and Se remains in \MoSeSe{} and \JanS.

The relaxed Janus MoSSe monolayer preserves its hexagonal symmetry, with a primitive lattice constant of 3.24~\AA, which corresponds to a $4\times4\times1$ supercell with in-plane parameters $a = 12.974$~\AA{} and $b = 12.980$~\AA{} and a lattice angle of $120.00^\circ$. This value is close to the lattice constant of $3.22 \pm 0.01$~\AA{} measured by low-energy electron diffraction for Janus MoSSe supported on Au(111) \cite{picker2025atomic}, and it is consistent with previous first-principles values of 3.22--3.25~\AA{} \cite{picker2025atomic, zhang2017janus}. We generated the two Janus-derived monolayers from this structure by removing the Se layer to obtain \JanSe or the S layer to obtain \JanS, and then fully relaxed them. Table~\ref{tab:lattice} summarizes the optimized lattice parameters.

\begin{table}[h!]
\centering
\caption{Optimized lattice parameters of the Janus MoSSe monolayer and the two depleted monolayers derived from it. The primitive lattice constant $a_0$ is the supercell parameter divided by four.}
\label{tab:lattice}
\setlength{\tabcolsep}{0pt}
\begin{tabular*}{\columnwidth}{@{\extracolsep{\fill}}lcccc}
\hline
Structure & $a$ (\AA) & $b$ (\AA) & $a_0$ (\AA) & $\gamma$ ($^\circ$) \\
\hline
MoSSe & 12.974 & 12.980 & 3.24 & 120.00 \\
\JanS & 12.371 & 12.373 & 3.09 & 120.05 \\
\JanSe & 11.824 & 11.843 & 2.96 & 120.01 \\
\hline
\end{tabular*}
\end{table}

Both depleted monolayers converge to locally optimized structures that preserve the parent phase's hexagonal symmetry, with lattice angles within $0.05^\circ$ of $120^\circ$. However, a pronounced contraction of the in-plane lattice is observed. The lattice parameters of \JanS{} are approximately 4.7\% smaller than those of the Janus parent, and those of \JanSe{} are approximately 8.8\% smaller. This contraction reflects the undercoordination of the Mo atoms after removing one chalcogen layer, which shortens the Mo-- Mo distances as the metal plane rebonds, analogous to that reported for sulfur-depleted MoS$_2$ \cite{jaques2026structural}. The smaller cell of \JanSe{} also follows from the smaller covalent radius of S and the correspondingly shorter Mo-S bonds.

The following uses two sets of simulations with different roles. The AIMD trajectories follow the two Janus-derived monolayers in a $4\times4$ cell with 32 atoms (16 chalcogen atoms) for 10~ps, either in the relaxed depleted cell of Table~\ref{tab:lattice} or with an in-plane barostat. The reactive simulations follow the four depleted monolayers and the depleted lateral MoS$_2$/MoSe$_2$ heterostructure of Figure~\ref{fig:systems}(b), in sheets of approximately 50 $\times$ 50~nm for 100~ps, in the box of the relaxed pristine parent. After the depletion, these sheets contain 57\,148 atoms for \MoSS, 52\,896 for \MoSeSe, and 54\,824 for the two Janus-derived sheets and for the lateral one. The two descriptions therefore differ in the sheet's mechanical state, as well as in size and simulation times.

\begin{figure*}[t!]
  \centering
  \includegraphics[width=16cm]{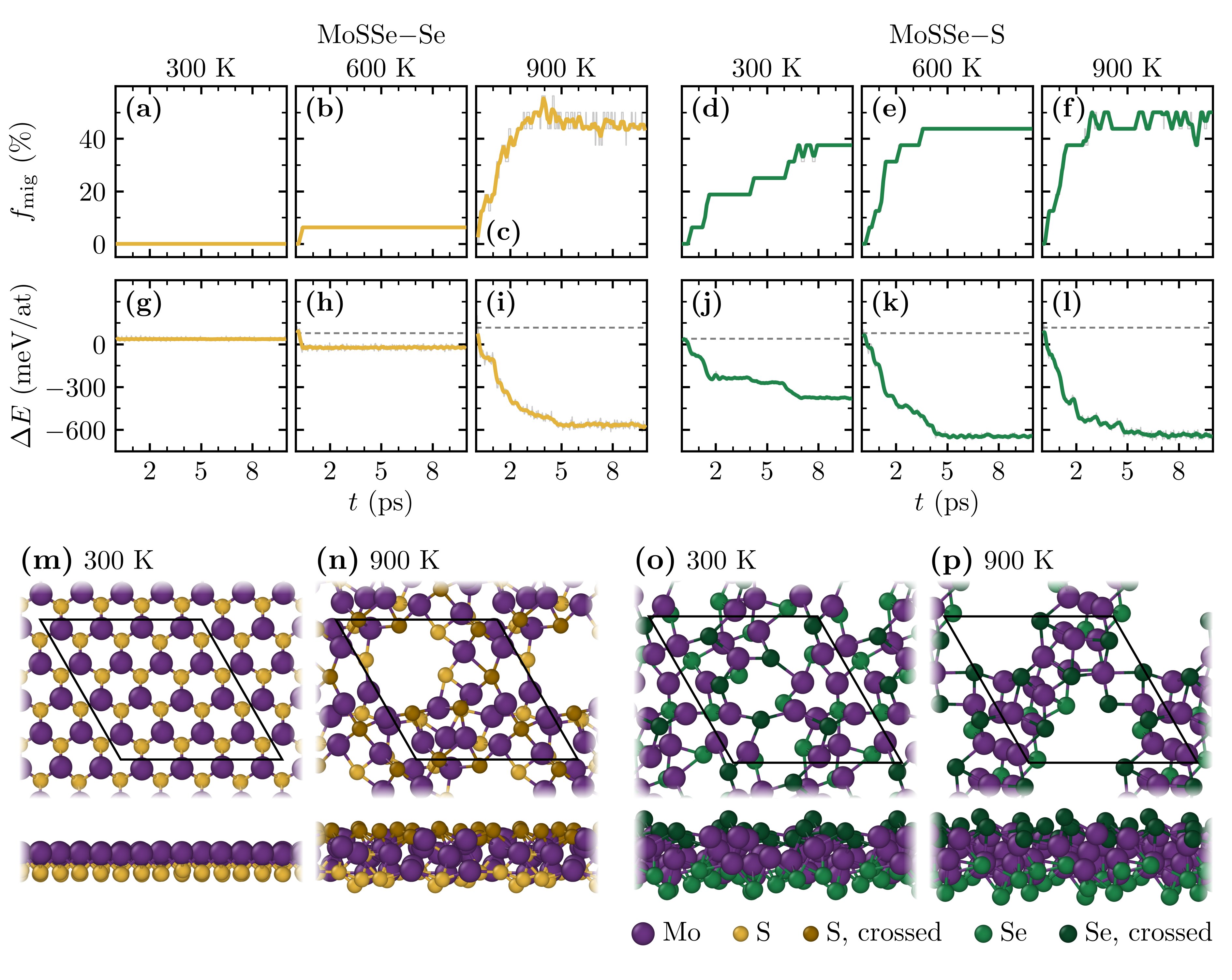}
  \caption{AIMD at fixed cell of \JanSe{} and \JanS{} at 300, 600, and 900~K. (a--f) Migrated fraction $\fmig$ and (g--l) Kohn-Sham energy per atom relative to $t = 0$, with the dashed line at $(3/2)k_{\mathrm{B}}T$. (m--p) Final structures at 300 and 900~K, with the atoms that crossed the Mo plane shown in a darker shade.}
  \label{fig:nvt}
\end{figure*}

The first question is how chalcogen atoms in each depleted monolayer leave/cross their structural layer, and what the energy cost of these processes is. To track structural reconstructions, we classify each chalcogen atom as migrated when it lies more than $\Delta Z = 0.5$~\AA{} beyond the Mo plane on the side opposite its initial layer, following the same criterion as in our previous work \cite{jaques2026structural}. The migrated fraction $\fmig$ is the fraction of chalcogen atoms in this condition. In the AIMD cell, the Mo plane is the mean out-of-plane coordinate of the Mo atoms. Figure~\ref {fig:nvt} presents the migrated fraction and the Kohn-Sham energy per atom relative to its initial value, $\Delta E$, obtained from AIMD at fixed cell at 300, 600, and 900~K.

When S remains, the sheet is preserved at low temperature. At 300~K, Figure~\ref{fig:nvt}(a) and (g) show that \JanSe{} maintains its initial atomic arrangement throughout the whole simulation, with the energy values fluctuating around a nearly constant value and no S atom crossing the Mo plane. The final energy change of $+36$~meV/atom corresponds to the thermal term $(3/2)k_{\mathrm{B}}T$ of 39~meV. At 600~K, a single S atom crosses during the initial stage of the trajectory ($\fmig = 6.25\%$), after which both the energy and the migrated fraction remain nearly constant, as shown in Figure~\ref{fig:nvt}(b) and (h). At 900~K, Figure~\ref{fig:nvt}(c) and (i) reveal a second and extended decrease in the energy after the initial relaxation, during which the migrated fraction rises steeply and saturates at approximately 45\%, with transient values that reach 56\%, and the energy change reaches $-567$~meV/atom.

In contrast, when Se remains, the sheet reconstructs already at room temperature. In \JanS{} at 300~K, Figure~\ref{fig:nvt}(d) and (j) show that the energy decreases through a sequence of well-defined steps, each followed by a temporary plateau, and each step is accompanied by an increase of $\fmig$ of $1/16 = 6.25\%$, that is, by the crossing of one Se atom. The migrated fraction reaches approximately 38\% (6 of 16 Se atoms) at the end of the trajectory, so that the Se atoms occupy both sides of the Mo plane, and the energy change is $-376$~meV/atom. At 600~K, the evolution in Figure~\ref{fig:nvt}(e) and (k) is considerably faster. The discrete events separated by large plateaus give way to an almost continuous decrease of the energy during the first 4~ps, while $\fmig$ rises to approximately 44\% within the same time window. At 900~K, Figures~\ref {fig:nvt}(f) and (l) show that the same pathway is followed more rapidly, with $\fmig$ saturating at 44--50\% within about 3~ps. The energy release is complete at 600~K, with $-646$~meV/atom, compared with $-639$~meV/atom at 900~K.

The energy released by reconstruction at a fixed cell reaches about 650~meV per atom, far above the thermal term, which ranges from 39--116~meV between 300 and 900~K. We also note that, at 900~K, the migrated fraction saturates within the first 2--3~ps, whereas the energy continues to decrease until approximately 5~ps. Crossing the Mo plane is therefore the fast step that triggers structural reconstruction. The subsequent relaxation of the energy is dominated by in-plane rearrangements of the two-sided structure, during which the mean distance from each Mo atom to its nearest Mo neighbor decreases from 2.7--2.9~\AA{} in the first 0.2~ps to 2.45--2.56~\AA{} in the last 2~ps, compared with 2.8--2.9~\AA{} in the runs that do not reconstruct. A similar shortening of the Mo-Mo distances at undercoordinated sites was reported for sulfur-depleted MoS$_2$ \cite{jaques2026structural}.

Counting crossings with a hysteresis of $\pm0.5$~\AA, we observe 6, 7, and 9 forward crossings of Se at 300, 600, and 900~K, and a single backward crossing at 900~K, whereas S shows 14 forward crossings at 900~K accompanied by 6 backward ones. With the free in-plane cell discussed below, however, the same counting gives 8 forward and no backward crossings for S at 900~K, and 10 forward and 3 backward ones for Se at 600~K. An atom that crosses and later stays within 0.5~\AA{} of the Mo plane is not counted as returned, so the net number of crossings can exceed the instantaneous $\fmig$ by one or two atoms. These counts come from single 10~ps trajectories with 16 chalcogen atoms, so we extract no rate or barrier from them, and we do not use event direction to distinguish between the two chalcogens.

Figure~\ref{fig:nvt}(m)--(p) shows the final structures obtained. The \JanS{} cell is already disordered at 300~K, whereas the \JanSe{} cell retains its hexagonal lattice at 300~K and loses it only at 900~K. The open/holey regions in the reconstructed cells result from large out-of-plane rearrangements in a cell of approximately 1.2~nm and are repeated by its periodic images, which limits what these snapshots can reveal about the extended sheet; we address this below with the reactive MD simulations. Within this limit, the fixed-cell trajectories already show that the remaining chalcogen affects how the sheet reconstructs and the temperature at which it occurs.

\begin{figure*}[t!]
  \centering
  \includegraphics[width=16cm]{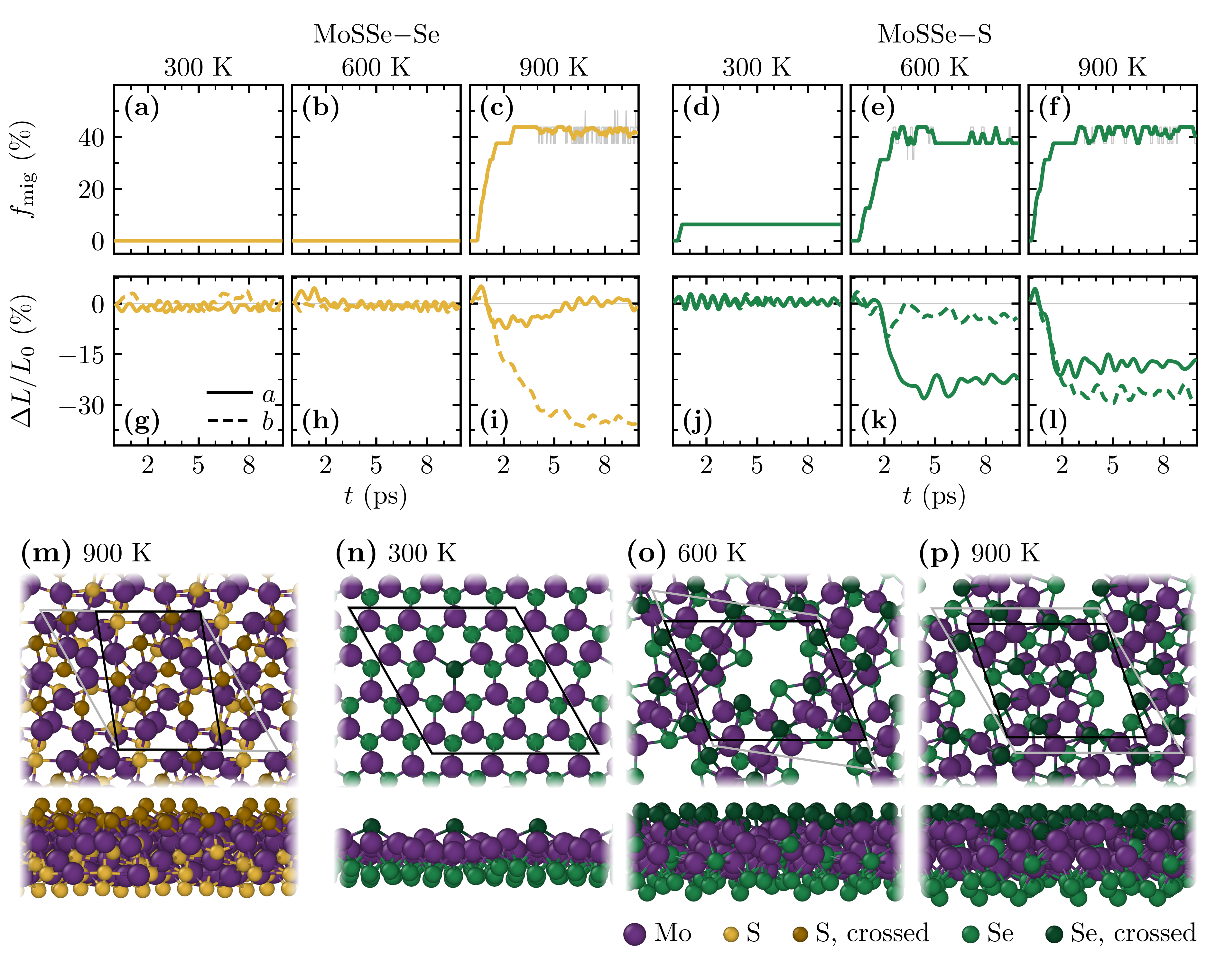}
  \caption{AIMD with an in-plane barostat of \JanSe{} and \JanS{} at 300, 600, and 900~K. (a--f) Migrated fraction $\fmig$ and (g--l) relative change of the in-plane cell lengths $a$ (solid) and $b$ (dashed). (m--p) Final structures of the runs in which chalcogen atoms cross the Mo plane, with the final cell in black and the initial cell in gray.}
  \label{fig:npt}
\end{figure*}

Because the fixed-cell simulations start from the relaxed depleted cell, they do not reveal how the lattice responds to reconstruction or whether residual in-plane stress contributes to it. To address these points, we repeated the AIMD simulations with a barostat acting only on the in-plane cell at zero target stress. Figure~\ref {fig:npt} presents the migrated fraction and the relative change in in-plane cell lengths.

In the regions where the sheet does not reconstruct, the free cell changes by approximately 1\%, and the lattice parameters fluctuate around a common average value with $a \approx b$, as shown in Figure~\ref{fig:npt}(g) and (h). The depleted cells used at fixed cell are therefore close to their stress-free state at finite temperature, and the reconstruction observed there does not require an applied in-plane strain.

At low temperatures, the rare events differ between the two simulation sets. In \JanS{} at 300~K, a single Se atom crosses at approximately 0.4~ps and no further events are observed, so $\fmig$ stays at 6.25\% in Figure~\ref{fig:npt}(d), compared with six crossings and approximately 38\% at fixed cell. In \JanSe{} at 600~K, the single crossing observed at fixed cell is absent, and $\fmig$ remains zero throughout the trajectory, as shown in Figure~\ref{fig:npt}(b). With one trajectory per condition, different thermostats, and a cell change of about 1\%, the present data do not allow us to separate the cell effect from variability between single trajectories. In our previous study, we assigned a primary role to in-plane strain in biasing S traversal in depleted MoS$_2$ \cite{jaques2026structural}, and the present trajectories neither confirm nor exclude this effect.

At higher temperatures, reconstruction proceeds in both simulation sets and reaches similar plateaus. For \JanSe{} at 900~K, the migration begins at approximately 0.6~ps with the nearly simultaneous crossing of two S atoms, followed by a sequence of individual events, and $\fmig$ saturates at approximately 42\%, with 6 to 7 of the 16 S atoms on the opposite side, within 2.5~ps, as seen in Figure~\ref{fig:npt}(c). Figure~\ref{fig:npt}(e) and (f) show that \JanS{} reconstructs at both 600 and 900~K, with the first crossings at approximately 0.6 and 0.3~ps, respectively, and migrated fractions that saturate at 38--42\%. Since these plateau values are close to those obtained at a fixed cell, the saturation level of the migration appears to be a property of the reconstructed two-sided structure, independent of the cell constraint.

The free cell reveals a feature that fixed-cell simulations cannot show: a strongly anisotropic in-plane contraction. For \JanSe{} at 900~K, Figure~\ref{fig:npt}(i) shows that the lattice parameter $b$ contracts from approximately 12 to 7.7~\AA, close to 35\%, while $a$ remains near its initial value. For \JanS, Figure~\ref{fig:npt}(k) and (l) show that one lattice vector contracts to about 9.5~\AA{} at 600~K while the other contracts by about 4\%, and that both reduce to about 10.1 and 9.1~\AA, respectively, at 900~K. The reconstruction thus breaks the cell's hexagonal symmetry, as the open regions created by out-of-plane rearrangements are partially closed by lattice contraction along preferential directions, as seen in the final structures in Figure~\ref{fig:npt}(m)--(p). The crossings come first, since the migrated fraction reaches half of its plateau (0.5--0.8~ps) before the cell reaches half of its contraction, indicating that the contraction follows reconstruction. Consistently, the reconstructed configurations reach lower energies with the free cell than at fixed cell, with an energy change of approximately $-0.8$~eV/atom ($-26$~eV per cell) against $-0.57$~eV/atom ($-18$~eV per cell) for \JanSe{} at 900~K, so that the relaxation of the cell provides an additional stabilization of the reconstructed sheet. The contraction magnitude is specific to a periodic cell of approximately 1.2~nm, and we interpret it as the reconstructed sheet's tendency to become denser.

\begin{figure*}[t!]
  \centering
  \includegraphics[width=16cm]{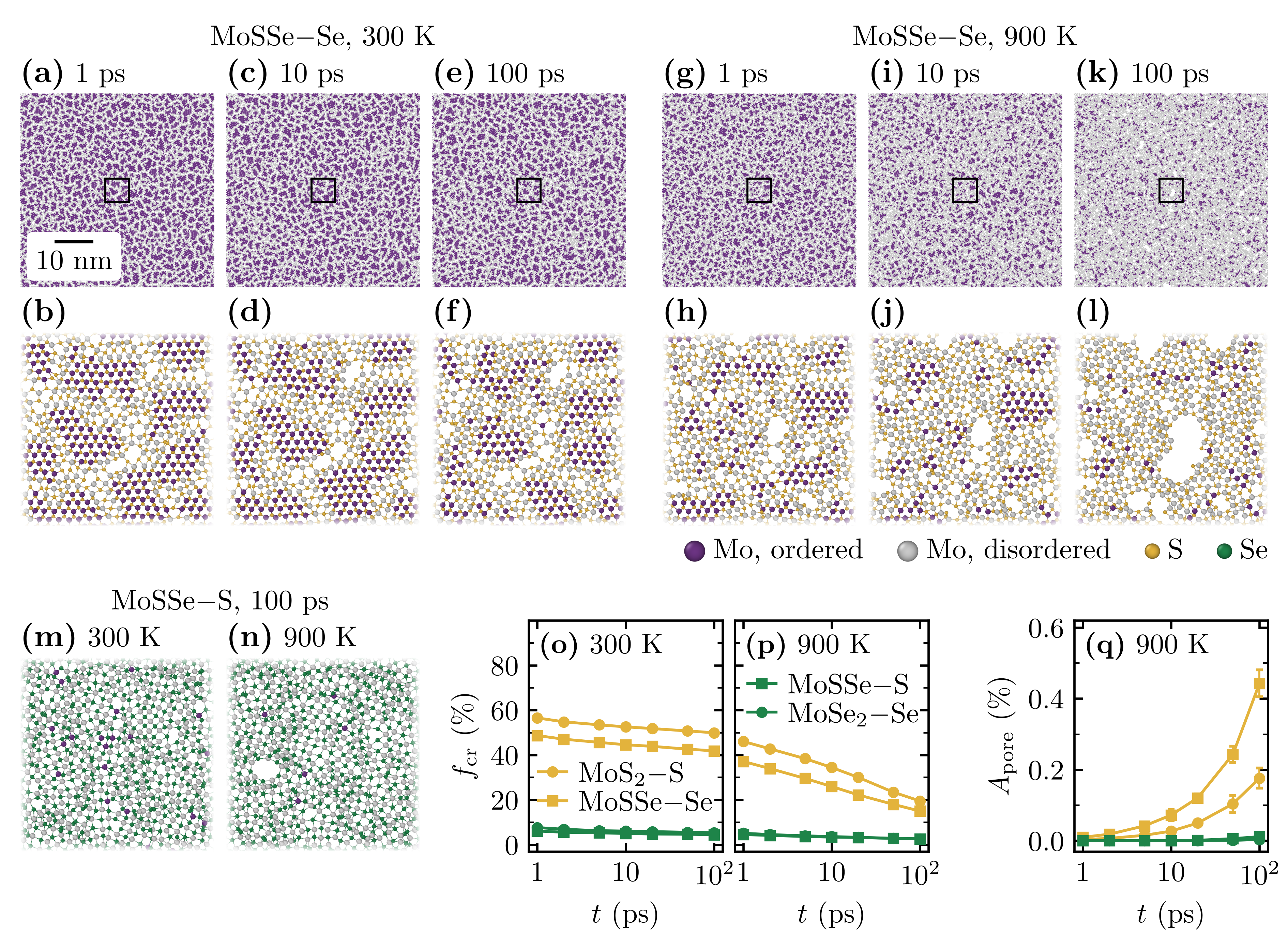}
  \caption{Reactive MD simulations of the 50 $\times$ 50~nm sheets, with ordered Mo atoms ($|\psi_6| > 0.7$) in purple and the others in gray. (a--l) \JanSe{} at 1, 10, and 100~ps, at 300~K (a--f) and 900~K (g--l), as the whole sheet (a, c, e, g, i, k) and as the 6 $\times$ 6~nm region marked by the square (b, d, f, h, j, l). (m, n) 6 $\times$ 6~nm regions of \JanS{} at 100~ps, at 300 and 900~K. (o, p) Ordered fraction $\fcr$ at 300 and 900~K and (q) pore area at 900~K against time.}
  \label{fig:domains}
\end{figure*}

The AIMD cell contains only 16 Mo atoms and their periodic images, so it cannot tell whether the reconstructed sheet retains any Mo lattice order beyond approximately 1.2~nm. To address this question, we have carried out reactive MD simulations of 50 $\times$ 50~nm sheets of the four depleted monolayers for 100~ps at 300, 600, and 900~K. The bond-orientational parameter measures the order of the Mo lattice \cite{halperin1978theory, nelson1979dislocation} of each Mo atom $j$, given by:

\begin{equation}
\psi_6(j) = \frac{1}{6}\sum_{k=1}^{6} e^{6i\theta_{jk}},
\label{eq:psi6}
\end{equation}

\noindent where the sum runs over the six nearest Mo neighbors $k$ of atom $j$ in the plane of the sheet and $\theta_{jk}$ is the angle of the $j$-$k$ bond with respect to a fixed in-plane axis. In a perfect hexagonal lattice, $|\psi_6| = 1$, and the value decreases as the six neighbors depart from the hexagonal arrangement. We count a Mo atom as ordered when $|\psi_6| > 0.7$, and the ordered fraction $\fcr$ is the fraction of Mo atoms that satisfy this condition. Since the extended sheets ripple, we take the Mo plane used for $\fmig$ as the local mean on a 20~\AA grid, and $\fmig$ changes by about 2 points or less with this choice of reference. Figure~\ref{fig:domains} shows the results.

Figure~\ref{fig:domains}(a)--(f) shows \JanSe, in which S remains, at 300~K. After 1~ps, ordered Mo atoms have already gathered into nanometric regions separated by disordered boundaries, both across the whole sheet and in the 6 $\times$ 6~nm region tracked over time. Inside these regions, the hexagonal Mo lattice and the S layer of the starting structure are preserved. The orientational correlation $g_6(r) = \mathrm{Re}\langle\psi_6(j)\psi_6^{*}(k)\rangle/\langle|\psi_6|^2\rangle$, averaged over the pairs of Mo atoms at a distance $r$ in the starting lattice and taken between 2.8 and 3.2~nm, remains at $+0.51$ in \JanSe{} and $+0.58$ in \MoSS{} at 300~K, which indicates that the ordered regions keep the orientation of the original lattice. They are therefore remnants of the original single crystal, with no sign of grain boundaries nucleated with independent orientations, and we refer to them as crystalline domains. These domains do not grow in time, as shown in Figure~\ref{fig:domains}(o), in which $\fcr$ of \JanSe{} decreases from 48.7\% at 1~ps to 41.8\% at 100~ps at 300~K. The decrease is considerably faster at 900~K, from 37.1\% to 15.0\%, as shown in Figure~\ref{fig:domains}(g)--(l) and (p). After 100~ps, the ordered fraction amounts to 50\%, 34\%, and 19\% in \MoSS{} and to 42\%, 26\%, and 15\% in \JanSe{} at 300, 600, and 900~K, respectively. The crystalline domains thus disappeared over time and with temperature, as expected when S atoms leave their layer via thermally activated events.

When Se remains, the Mo lattice is disordered from the first stored frame. In \JanS, $\fcr$ is only 6.1\% at 1~ps and 4.5\% at 100~ps at 300~K, and 4.5\% and 2.6\% at 900~K, as seen in Figure~\ref{fig:domains}(m)--(p), while the orientational correlation at 3~nm vanishes. We also obtain values between 2.6\% and 5.3\% after 100~ps for \MoSeSe. Temperature and time, which control the ordered fraction of sheets in which S remains, change it here by only a few points, since crystalline order is essentially lost before the first picosecond. The comparison between the two parents of each chalcogen shows that the parent lattice has a minor role. \MoSS{} and \JanSe{} follow similar curves, separated by 5--9 points in $\fcr$, and \MoSeSe{} and \JanS{} are nearly indistinguishable, so the remaining chalcogen atoms set the reconstruction mode. This picture is consistent with the final AIMD structures of Figure~\ref{fig:nvt}(m)--(p). The present sheets have lost half their chalcogen atoms and are followed for 100~ps, a regime far from that of the Se-deficient MoSe$_2$ monolayers studied experimentally, where a lower Se deficiency is accommodated by ordered mirror twin boundaries and inversion domains \cite{lehtinen2015atomic, lin2015vacancy}. In both cases, the Mo sublattice itself rearranges when Se is missing, and the loss of hexagonal order reported here is the short-time limit of this tendency when a whole layer is removed. Similarly, monolayer MoS$_2$ can be amorphized under ion irradiation \cite{valerius2020reversible}, so that the persistence of crystalline domains when S remains is constrained to the conditions examined here.

The two modes also differ in the defects they produce, although these differences are small. Pores, identified as the regions of the sheet with no atom within 3~\AA{} on a sampling grid of 1~\AA, are observed essentially only when S remains and at 900~K, and they grow with time, as shown in Figure~\ref{fig:domains}(q). In \JanSe, the pore area is negligible at 1~ps and reaches 0.44\% of the sheet at 100~ps, with the largest pore covering approximately 160~\AA$^2$, whereas the sheets in which Se remains stay essentially free of pores. Given the 100~ps window and the fixed box of the simulations, we regard this result as a trend, and no pore density is inferred from it. In addition, taking the Mo plane from a local fit within 6~\AA{} and counting as displaced the chalcogen atoms that lie less than 0.5~\AA{} from this plane or beyond it, the displaced S atoms at 300~K are weakly clustered, with a Warren-Cowley parameter \cite{cowley1950approximate}, computed on the neighbor graph of the starting lattice and equal to zero for a random arrangement, of $+0.11$ to $+0.15$ that vanishes at 900~K, while the displaced Se atoms are weakly anticorrelated, with values between $-0.07$ and $-0.12$. In the sheets in which S remains, the local fraction of displaced atoms at 300~K is also correlated with the local height of the Mo plane, with correlation coefficients between $-0.6$ and $-0.7$, so that the departure of S from its layer is accompanied by a rippling of the initially flat sheet.

We use the reactive model here for sheet morphology and its trends with chalcogen, time, and temperature, while we take bond lengths and energies from the AIMD. The first-neighbor Mo-S and Mo-Se peaks are approximately 0.3~\AA{} longer than in the AIMD (2.57--2.62~\AA{} against 2.22--2.32~\AA{} for Mo-S and 2.87--2.97~\AA{} against 2.57--2.67~\AA{} for Mo-Se). The distance from each Mo atom to its nearest Mo neighbor also decreases upon reconstruction in the reactive sheets, from 3.18--3.27~\AA{} in the starting lattice to 2.89--3.05~\AA{} after 100~ps. The longer distances in the reactive model are attributed to the force field and, in part, to the different mechanical conditions of the two simulation sets, since the AIMD starts from the relaxed depleted cell. In contrast, the reactive sheets are kept in the box of the pristine parent.

\begin{figure*}[t!]
  \centering
  \includegraphics[width=16cm]{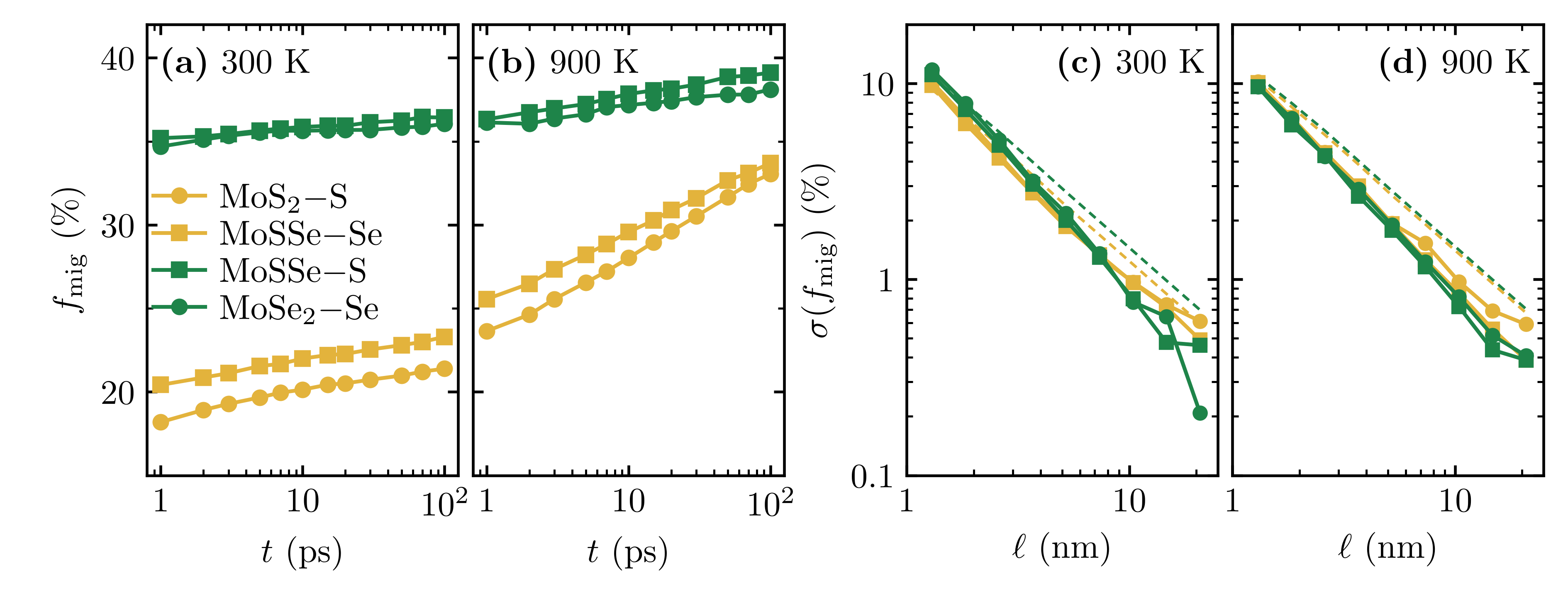}
  \caption{Time and length scales in the reactive MD simulations of the four depleted monolayers. (a, b) Migrated fraction $\fmig$ against time at 300 and 900~K. (c, d) Standard deviation of $\fmig$ among square windows of side $\ell$ at 100~ps, at 300 and 900~K, with the binomial expectation as dashed lines.}
  \label{fig:scales}
\end{figure*}

The AIMD trajectories follow 16 chalcogen atoms for 10~ps, raising two questions about how far these conditions limit what they can reveal: whether migration finishes within that time and how much a cell of that size fluctuates around the mean of an extended sheet. The reactive simulations answer both, since they cover 100~ps and approximately 27\,000 chalcogen atoms per sheet. Figure~\ref{fig:scales} presents the corresponding analysis.

When Se remains, the migration is essentially complete at 1~ps, the first stored frame, and at most 3.3\% of the final value is added after 10~ps at any temperature, as shown in Figure~\ref{fig:scales}(a) and (b), so that 10~ps covers the whole process. When S remains, we observe two stages: a fast one within the first picosecond and a slow one that is still in progress at 100~ps. In \MoSS{} at 900~K, $\fmig$ increases from 24\% at 1~ps to 33\% at 100~ps, and in both sheets, S remains 12--15\% of the final value at 900~K only after the 10~ps that an AIMD trajectory covers. After 100~ps, $\fmig$ amounts to 21.4\%, 28.0\%, and 33.1\% in \MoSS{} and to 23.3\%, 29.3\%, and 33.7\% in \JanSe{} at 300, 600, and 900~K, respectively, whereas it remains at 36.1\%, 37.2\%, and 38.1\% in \MoSeSe{} and at 36.4\%, 38.1\%, and 39.1\% in \JanS. Thus, S migration increases by about 10--12 points between 300 and 900~K, while Se changes by 2--3 points, with a standard deviation of 0.4 points or less among the three seeds. The two parent structures of each chalcogen give the same $\fmig$ within approximately 2 points, which is the sensitivity of $\fmig$ to the definition of the local Mo plane. The absolute time scales of a reactive force field should be taken as trends, and the conclusions rest on the dependence on the chalcogen and on temperature.

The effect of the cell size is addressed in Figure~\ref{fig:scales}(c) and (d), which show the standard deviation of $\fmig$ among square windows of side $\ell$ cut from the sheets at 100~ps. For windows of 1.3~nm, comparable to the AIMD cell of about 1.2~nm, $\fmig$ fluctuates by about 10 points around the sheet mean. The deviation decreases approximately as $1/\ell$, close to the binomial expectation for uncorrelated atoms when S remains and somewhat faster when Se remains, and falls below approximately 2 points for windows larger than 5~nm. In addition, in a $4\times4$ cell $\fmig$ can only change in steps of 1/16, that is, 6.25 points. Differences of a few points between two AIMD trajectories, or between an AIMD trajectory and an extended sheet, are therefore within the sampling noise of the small cell.

However, this noise does not account for the main difference between the two descriptions. At 300~K, no S atom crosses the Mo plane in the AIMD of \JanSe, whereas 21--23\% of S atoms cross after 100~ps in the reactive sheets. According to the window statistics, a 1.3~nm window with no migrated atom occurs in only 2--4\% of positions. The crystalline domains of Figure~\ref{fig:domains} contribute in the same direction, since a 1.3~nm window has at least 80\% of its Mo atoms ordered in 2--8\% of the positions at 300~K, which is more than ten times the value expected for randomly placed ordered atoms, but this remains a small effect. Calculated diffusion barriers above 1.9~eV have been reported for isolated S vacancies in MoS$_2$ \cite{sensoy2017strain}. However, they refer to a different event than crossing the Mo plane in a sheet with one empty side, and also align with the absence of S crossings at 300~K in the AIMD. The remaining difference is attributed to the force field and, possibly, to the mechanical state of the sheets. The reactive sheets are kept in the box of the pristine parent structure, where those with remaining S start under tension, as indicated by the negative simulation-box pressure at the start of the dynamics.

In contrast, the AIMD starts from the relaxed, depleted cell. As discussed below for the lateral heterostructure, partial relief of this tension does not reduce S migration, so we do not assign the difference to the stress state alone. We therefore use reactive simulations to determine the morphology and scale of the two reconstruction modes, and we take the temperature at which S migration sets in from the AIMD.

\begin{figure*}[t!]
  \centering
  \includegraphics[width=16cm]{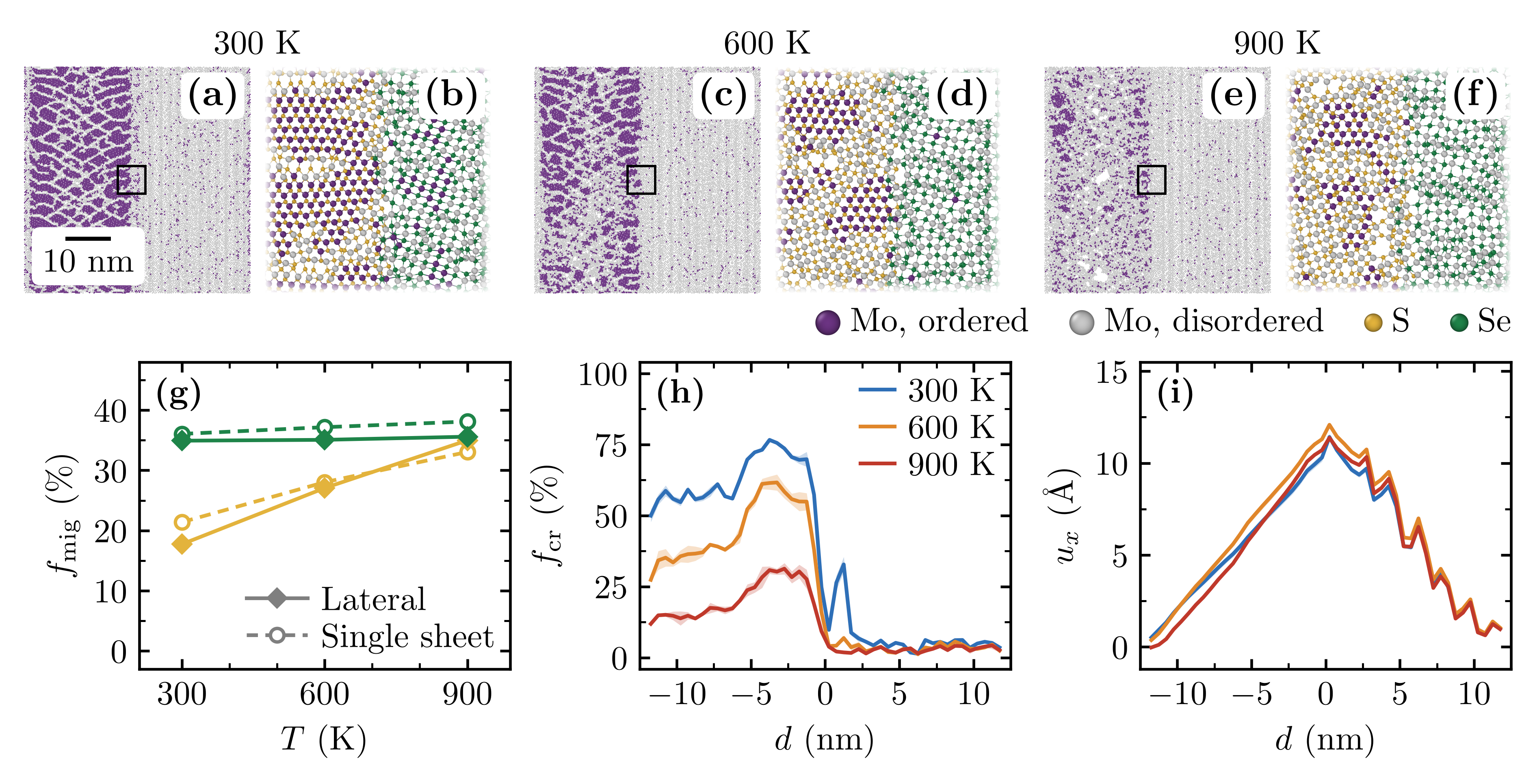}
  \caption{Depleted lateral heterostructure after 100~ps, with \MoSS{} on the left and \MoSeSe{} on the right. (a--f) Whole sheet (a, c, e) and 6 $\times$ 6~nm region at the interface (b, d, f) at 300, 600, and 900~K. (g) Migrated fraction of each chalcogen against temperature in the lateral sheet and in the single sheets. (h) Ordered fraction $\fcr$ and (i) collective displacement $u_x$ of the Mo lattice against the initial distance $d$ to the interface.
}
  \label{fig:lateral}
\end{figure*}

So far, the comparisons involve separate sheets, each with its own box and thermal history. A more demanding test of the chalcogen's role is to place S and Se in the same sheet, on one Mo lattice and at one temperature. For this purpose, we have simulated a lateral MoS$_2$/MoSe$_2$ heterostructure of 49.5 $\times$ 49.5~nm from which the same chalcogen layer was removed on both sides, as illustrated in Figure~\ref{fig:systems}(b), so that one half is \MoSS{} and the other is \MoSeSe. Under periodic boundary conditions, the sheet contains two interfaces, and $d$ denotes the initial distance from an atom to the nearest one, negative in the \MoSS{} half. Figure~\ref{fig:lateral} shows the results after 100~ps.

Figure~\ref{fig:lateral}(a)--(f) shows that the two reconstruction modes coexist, each confined to its half. The S half retains crystalline domains that are consumed as temperature increases, while the Se half remains disordered at all temperatures. The corresponding fractions are close to those of the single sheets, as shown in Figure~\ref{fig:lateral}(g). At 300, 600, and 900~K, $\fmig$ of S is 17.7\%, 27.1\%, and 35.0\% in the lateral sheet against 21.4\%, 28.0\%, and 33.1\% in \MoSS, and $\fmig$ of Se is 34.9\%, 35.1\%, and 35.6\% against 36.1\%, 37.2\%, and 38.1\% in \MoSeSe. The ordered fraction of the Se half, 7.0\%, 3.6\%, and 2.9\%, also follows that of \MoSeSe{} (5.3\%, 3.6\%, and 2.6\%). The S half is more ordered than \MoSS{} at 300 and 600~K, with $\fcr$ of 60.9\% and 43.0\% against 49.9\% and 33.9\%, and the two coincide at 900~K (20.1\% against 19.5\%). The profile in Figure~\ref{fig:lateral}(h) shows that part of this difference originates in a zone of about 5~nm on the S side of the interface, where $\fcr$ reaches 70--77\% at 300~K, versus about 57\% farther inside. In the same zone, only 12--17\% of S atoms migrate, compared with 21\%. Away from this zone, S migration at 300~K is the same as in the single sheet (21.2\% vs 21.4\%), and at 600 and 900~K it is slightly higher (30.7\% vs 28.0\% and 37.6\% vs 33.1\%).

The two halves also exchange strain, as revealed by $u_x$ in Figure~\ref{fig:lateral}(i), the mean displacement of the Mo atoms along the direction normal to the interface, $x$, with respect to the starting lattice, averaged in bins of $d$ and taken as positive toward the \MoSS{} half. This is a collective lattice displacement, since no Mo atom changes site. The interface moves by about 1~nm into the S half within the first 10~ps and by 1.1--1.2~nm after 100~ps, so the S half contracts by about 9\% along $x$ and the Se half stretches by the same amount. This contraction is comparable in magnitude to that of the relaxed cells obtained from first principles (Table~\ref{tab:lattice}), and it is consistent with the tensile state of the single sheets in which S remains when kept in the box of its pristine parent. The lateral sheet therefore tests the role of this tension. The contraction of approximately 9\% of the S half along $x$ preserves more of its crystalline order but does not reduce S migration, as indicated by the values above. Although the relief is uniaxial, because the box remains fixed along the interface, this result suggests that tension in the fixed box is not the sole origin of S migration at 300~K in the reactive description.

The lateral sheet also clarifies that Se migration requires no thermal activation. The energy minimization that precedes the dynamics, which involves no thermal energy, already displaces 34\% of the Se atoms across the Mo plane, the same fraction found between 300 and 900~K, against 9\% of the S atoms. This displacement extends over the whole Se half but is not uniform, since in strips of 0.5~nm parallel to the interface the migrated fraction of Se ranges from 1\% to 69\%. The Se half remains banded during the dynamics, although the bands rearrange within the first 5--10~ps. The single sheets, in turn, do not change during the minimization, since nothing perturbs their symmetric starting configuration. The migration of Se thus proceeds essentially without thermal activation once the symmetry of the starting configuration is broken, here by the interface and in the single sheets by the first thermal displacements. In contrast, most S migration requires temperature. This minimization stopped at the energy criterion, with residual forces that were still large, and it is reported as such, without any claim to the sheet's lowest-energy configuration.

These results indicate that the reconstruction mode is a local property of the sheet, determined by the chalcogen on each side of the interface, and that it is preserved a few nanometers away. From this perspective, the chalcogen pattern of the parent monolayer, which can be defined by the epitaxial growth of lateral MoS$_2$/MoSe$_2$ heterostructures \cite{chen2019lateral} or by patterned chalcogen replacement \cite{mahjouri2015patterned, taghinejad2020synthetic}, and which is also of interest for photocatalysis \cite{zhao2025rational}, sets where a depleted sheet remains crystalline and where it becomes disordered.

\section{Conclusions}

In summary, we have combined \textit{ab initio} and reactive molecular dynamics simulations to investigate the spontaneous structural reconstructions of the monolayers obtained by removing one chalcogen layer from MoS$_2$, MoSe$_2$, and Janus MoSSe. The results show that the remaining chalcogen determines the reconstruction mode, with only a minor influence from the parent lattice. When Se remains, it crosses the Mo plane already at 300~K in the \textit{ab initio} trajectories, and the reconstruction releases up to approximately 650~meV/atom at the fixed cell. When S remains, no atom crosses at 300~K, and the sheet reconstructs only at 900~K. With a free in-plane cell, the crossings precede the lattice response, indicating that anisotropic cell contraction results from structural reconstruction.

At the 50~nm scale, the two modes produce different sheets. With S, the Mo lattice remains crystalline in nanometric domains that disappear with time and thermal heating, and pores/holes appear at 900~K. With Se, the Mo lattice loses its crystalline order within the first picosecond, and the migrated atomic fraction hardly changes between 300 and 900~K. The reactive simulations also set the limits of the small cell, since part of the S migration at 900~K occurs after 10~ps and the migrated fraction in a region the size of the AIMD cell fluctuates by about 10 points. While the reactive description places the onset of S migration at a lower temperature than the first-principles one, they show similar trends.

The depleted lateral MoS$_2$/MoSe$_2$ heterostructure shows that the two modes coexist in one sheet, each half reconstructing like the corresponding single sheet, with similar migration fractions, while the S half contracts at the expense of the Se half. These findings suggest that the composition of the parent monolayer, in Janus or laterally patterned form, can define which regions of a chalcogen-deficient sheet retain a crystalline Mo lattice and which become disordered, with possible implications for the thermal processing of these materials and for generating undercoordinated Mo sites.

\section{Methods}

We performed first-principles simulations within density functional theory (DFT) using the SIESTA package \cite{soler2002siesta, garcia2020siesta}. We described exchange-correlation effects within the generalized gradient approximation using the Perdew-Burke-Ernzerhof (PBE) functional \cite{perdew1996generalized}. We replaced core electrons with norm-conserving pseudopotentials \cite{hamann2013optimized} from the PseudoDojo library \cite{vansetten2018pseudodojo} in the PSML format \cite{garcia2018psml}, and used a double-$\zeta$ polarized basis set for valence electrons \cite{junquera2001numerical} with an energy shift of 0.01~Ry and a split norm of 0.15. We defined the real-space integration grid with a mesh cutoff of 350~Ry. The initial structure consisted of a $4\times4\times1$ Janus MoSSe supercell containing 48 atoms, with a 20~\AA{} cell along the out-of-plane direction, leaving a vacuum region of 17--19~\AA. We performed structural optimizations using the conjugate-gradient algorithm with variable in-plane lattice parameters and a fixed out-of-plane lattice vector until the residual forces fell below 0.02~eV/\AA. We performed Brillouin-zone integrations using a $2\times2\times1$ Monkhorst-Pack mesh \cite{monkhorst1976special}, and achieved self-consistency with a density matrix tolerance of $10^{-4}$ and a mixing weight of 0.02. We obtained the \JanSe{} and \JanS{} structures, each containing 32 atoms, by removing the Se and S layers, respectively, from the optimized Janus geometry, and fully relaxed them with the same parameters. All fixed-cell AIMD simulations start from these relaxed depleted cells.
The calculations were not spin-polarized and used an electronic temperature of 0.01~eV and scalar-relativistic pseudopotentials.

We performed AIMD simulations at 300, 600, and 900~K with a 1~fs time step and a total duration of 10~ps, using $\Gamma$-point sampling of the Brillouin zone. The fixed-cell (NVT) simulations used a Nos\'e thermostat \cite{nose1984unified, nose1984molecular} with a thermostat mass of 30~Ry$\cdot$fs$^{2}$. We carried out the simulations with a free in-plane cell (NPT) by coupling SIESTA, acting as the force engine, to the Atomic Simulation Environment (ASE) \cite{larsen2017atomic} through the i-PI socket protocol \cite{ceriotti2014pi}. This setup allows the barostat to act exclusively on the in-plane degrees of freedom. The combined Nos\'e-Hoover and Parrinello-Rahman dynamics implemented in ASE \cite{melchionna1993hoover, melchionna2000constrained, holian1990time, parrinello1981polymorphic} was used with a strain mask that restricts the cell fluctuations to the $xx$, $yy$, and $xy$ components, so that the out-of-plane lattice vector, and hence the vacuum spacing, is kept fixed. We set the target in-plane stress to zero, with thermostat and barostat characteristic times of 25 and 75~fs, respectively, and defined the barostat fictitious mass from a reference bulk modulus of 100~GPa. We drew initial velocities from a Maxwell-Boltzmann distribution at the target temperature and removed the center-of-mass linear and angular momenta.

Reactive molecular dynamics simulations were performed with LAMMPS \cite{thompson2022lammps} using the ReaxFF formalism \cite{vanduin2001reaxff, chenoweth2008reaxff, aktulga2012parallel}, with the parameter set of Nayir \textit{et al.} \cite{nayir2020modeling}, the same employed in our previous study of sulfur-depleted MoS$_2$ \cite{jaques2026structural}. This set describes Mo, S, and Se together, with Mo-S bond and angle terms that coincide with those of Ostadhossein \textit{et al.} \cite{ostadhossein2017reaxff}. Its S-Se cross terms act only along the interface lines of the lateral heterostructure, since each sheet retains a single chalcogen species.
Atomic charges were equilibrated at every step \cite{rappe1991charge, nakano1997parallel} with a tolerance of $10^{-6}$. We built orthorhombic supercells of approximately 50 $\times$ 50~nm, with approximately 30~nm of vacuum along the out-of-plane direction, for MoS$_2$, MoSe$_2$, Janus MoSSe, and a lateral MoS$_2$/MoSe$_2$ heterostructure. The lateral sheet was built with a common lattice parameter, the mean of those of MoS$_2$ and MoSe$_2$, with the interface normal to the $x$ direction, and it contains two interfaces because of the periodic boundary conditions. We relaxed the pristine sheets by conjugate-gradient minimization with the in-plane box free. We then removed one chalcogen layer and minimized the atomic positions with a fixed box, with a relative energy tolerance of $10^{-8}$ and a force tolerance of $10^{-10}$~kcal~mol$^{-1}$~\AA$^{-1}$. We ran the dynamics in a fixed box with a Nos\'e-Hoover thermostat \cite{shinoda2004rapid, martyna1994constant, tuckerman2006liouville} with a damping time of 50~fs and a time step of 0.5~fs for 100~ps at 300, 600, and 900~K, storing configurations every 1~ps and using three independent sets of initial velocities for each condition. We rendered the structures with OVITO \cite{stukowski2010visualization}.

\section{Data Availability}

The data that support the findings of this study are available from the corresponding authors upon reasonable request.

\section{Author Contributions}

B.I., M.L.P.J. and R.B.O performed the calculations, with the assistance of G.S.L.F. M.L.P.J. prepared the figures. P.M.A. contributed to the conception of the study and to funding acquisition. M.L.P.J. and D.S.G. conceived and supervised the project. The manuscript was written through contributions of all authors. All authors have given approval to the final version of the manuscript.

\section{Notes}

The authors declare no competing financial interest.

\begin{acknowledgement}
\noindent B.I. thanks CNPq (process 194155/2025-0) and FAPESP (process 2024/11016-0). G.S.L.F. thanks FAPESP (process 2024/03413-9). R.B.O. thanks CNPq under process \#151043/2024-8. M.L.P.J. acknowledges financial support from FAPDF (grant 00193-00001807/2023-16), CNPq (grants 444921/2024-9 and 308222/2025-3), and CAPES (grant 88887.005164/2024-00). D.S.G. acknowledges the Center for Computing in Engineering and Sciences at Unicamp for financial support through the FAPESP/CEPID grant 2013/08293-7. The authors thank the Coaraci Supercomputer for computer time (FAPESP grant 2019/17874-0). D. S. G. acknowledges support from
INEO/CNPq and FAPESP (grant 2025/27044-5).
\end{acknowledgement}

\bibliography{references}

\end{document}